\documentclass[]{aa}
\usepackage{graphicx}
\begin{document}

\title{An empirical temperature calibration for the $\Delta a$ 
photometric system. I. The B-type stars\thanks{Based on observations at 
the Leopold-Figl Observatory for Astrophysics, University of Vienna}}
\author{E.~Paunzen\inst{1}, A.~Schnell\inst{1}, H.M.~Maitzen\inst{1}}

\mail{Ernst.Paunzen@univie.ac.at}

\institute{Institut f{\"u}r Astronomie der Universit{\"a}t Wien,
           T{\"u}rkenschanzstr. 17, A-1180 Wien, Austria}

\date{Received 1 June 2005 / Accepted 21 August 2005}
\titlerunning{An empirical temperature calibration for the $\Delta a$: B-type stars}{}

\abstract{We establish an empirical effective temperature calibration of
main sequence, luminosity class V to III B-type stars for the $\Delta a$ 
photometric system which was originally developed to
detect magnetic chemically peculiar objects of the upper
main sequence (early B-type to early F-type) at 5200\,\AA. 
However, this system provides the index $(g_1-y)$ which
shows an excellent correlation with $(B-V)$ 
as well as $(b-y)$ and can be used as an indicator of the effective temperature.
This is supplemented by a very accurate color-magnitude
diagram, $y$ or $V$ versus $(g_1-y)$, which can be used, for example, 
to determine the reddening, distance and age of an open cluster.
This makes the $\Delta a$ photometric system an excellent tool to
investigate the Hertzsprung-Russell-Diagram (HRD) in more detail.
Using the reddening-free parameters and already established calibrations within
the Str{\"o}mgren $uvby\beta$, Geneva 7-color and Johnson $UBV$ systems,
a polynomial fit of third degree for the averaged effective temperatures to 
the individual $(g_1-y)_0$ values was derived. For this purpose, data 
from the literature as well as new observations were taken resulting 
in 225 suitable bright normal B-type objects. The statistical
mean of the error for this sample is 238\,K which is 
sufficient to investigate the HRD of distant galactic 
open clusters as well as extragalactic aggregates in the future.
}
\maketitle

\section{Introduction}

One of the most important observational diagnostic tools of astrophysics
is the Hertzsprung-Russell-Diagram (HRD) allowing the study of the correlation
between the effective temperature and the absolute magnitude (or luminosity)
of astronomical objects. Virtually all stellar 
astrophysical models are tested according to the HRD.

The absolute magnitudes of stellar objects can be derived directly via
parallax measurements, appropriate photometric indices (e.g. Str{\"o}mgren
$\beta$) or on a statistical basis in open and globular clusters. The errors
of such combined estimates are already rather small ($<$\,0.1\,mag). 

The photometric calibration of effective temperatures is still a very
tricky business with several pit falls. The applied calibrations 
depend on the investigated spectral range and the physics introduced
in the models. Smalley \& Kupka (1997) give an excellent overview of how
convection, for example, can lead to significant deviations for A-type
and cooler stars. The statistical calibration of effective temperatures
via photometric indices has been done since the introduction of photometric
systems (Johnson 1958, Str{\"o}mgren 1966 and references therein). The
applied calibrations have become
more precise and sophisticated as the theoretical stellar atmospheres and the
input physics became more realistic. Furthermore, the amount of available photometric
and spectroscopic data is increasing. 

Almost thirty years ago, Maitzen (1976) introduced the narrow-band, three filter 
$\Delta a$ photometric system in order to investigate the flux depression at 5200\AA\,
which is very likely enhanced by the effects of magnetic radiative transport phenomena
within magnetic chemically peculiar stars of the upper main sequence (Kupka et al.
2003). Since then, more than two dozen papers have presented photoelectric and CCD 
results for galactic field stars, open clusters as well as the Large Magellanic Cloud
in the relevant spectral range from early B to early F-type 
(Maitzen 1993, Paunzen et al. 2003, Paunzen et al. 2005a,b).
The $\Delta a$ system is based on the three narrow band filters $g_1, g_2$ and $y$
from which two indices are calculated $(g_1-y)$ and $a$\,=\,$g_2 - (g_1 + y)/2$.
An $a$ versus $(g_1-y)$ diagram can easily detect magnetic chemically peculiar stars
(Sect. \ref{additional})
whereas the $(g_1 - y)$ index shows an excellent correlation with $(B-V)$ 
as well as $(b-y)$ and can be used as an indicator of the effective temperature.
The main result of detecting peculiar stars is supplemented by a very accurate 
color-magnitude
diagram, $y$ or $V$ versus $(g_1-y)$, which can be used, for example, 
to determine the reddening, distance and age of an open cluster
(Claret et al. 2003).

In the last few years, much theoretical effort was made to explain the results of
the $\Delta a$ photometric system in astrophysical terms. Kupka et al. (2003, 2004)
established a synthetic photometric $\Delta a$ system and confirm the observed dependency 
of the $a$ index as a function of various colour indices sensitive to 
the effective temperature and surface gravity variations within the Str{\"o}mgren
$uvby\beta$ and Johnson $UBV$ photometric systems using fluxes from ATLAS9
model atmospheres as well as most recent atomic line data together with 
opacity distribution function for individual chemical compositions.  Furthermore,
Claret et al. (2003) calculated isochrones, taking into
account mass loss during the main sequence evolution, for the $\Delta a$ photometric system
on the basis of modern equations of state including partial ionization through Saha's
approximation, the pressure of gas and radiation as well as the equations for
degenerate electrons. 

We now propose to establish an empirical effective temperature calibration for
main sequence (luminosity class V to III) stars in terms of $(g_1-y)_0$. This is
most important in studying very distant galactic open clusters and extragalactic systems
for which, in general, no photometric data within a standard system are available.
The absolute magnitudes and thus luminosities can be easily estimated via $y$ and
the appropriate isochrones. In this first of two papers we deal with B-type stars
because : 1) There are reddening-free parameters available allowing
us to check the dereddening procedure for
$(g_1-y)$ and 2) The stellar interiors for these objects are 
very similar over the whole spectral range. The second paper will
deal with A-type to early F-type objects for which the calibration has to be more
sophisticated due to the increase of line blanketing and luminosity effects.

For the empirical temperature calibration, we use a homogeneous
sample of bright ($V$\,$<$\,7\,mag), apparently normal type objects from the 
literature. But we also
present new observations for 99 B-type objects that are included in our
analysis. In total, 225 stars were used to derive effective temperatures within
the Str{\"o}mgren $uvby\beta$, Geneva 7-color and Johnson $UBV$ systems which
were then applied to establish a calibration in terms of $(g_1-y)_0$. The final
calibration is valid for effective temperatures between 33000 and 10000\,K
and yields a statistical mean error of 238\,K for the whole spectral range.

\begin{figure}
\begin{center}
\includegraphics[width=70mm]{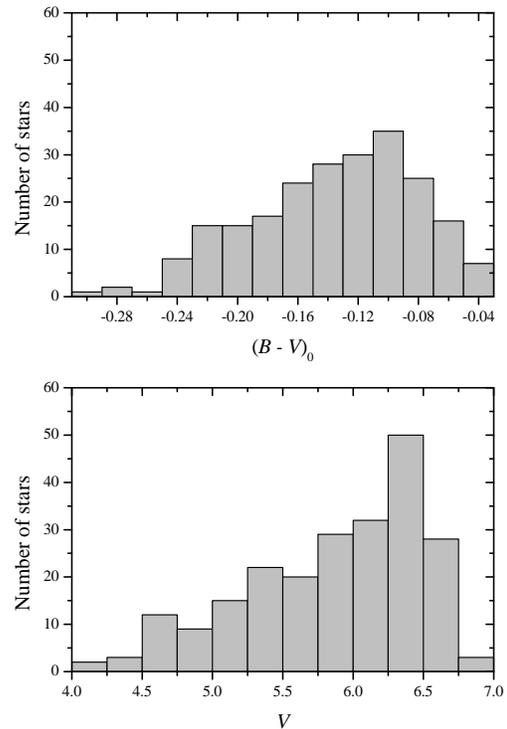}
\caption{The distribution of Johnson $V$ and $(B-V)_0$, see 
Sect. \ref{reddening}, for
our sample of 225 main sequence B-type objects.}
\label{hists}
\end{center}
\end{figure}

\section{Sample of program stars} \label{sops}

We have selected ``normal'' B-type objects fulfilling the
following criteria:
\begin{itemize}
\item classification as B-type, luminosity class V to III
\item no significantly deviating $\Delta a$ values, i.e. $\pm$10\,mmag 
\item not listed in the catalogue of Ap and Am stars by 
Renson et al. (1991)  
\item available data within the Johnson, Str{\"o}mgren,
Geneva and $\Delta a$ photometric system 
\end{itemize}
Binary systems of all kinds and high $v\sin i$ stars were a-priori not excluded.
The Johnson, Geneva and Str{\"o}mgren colors were taken from the General Catalogue of 
Photometric Data (GCPD, Mermilliod et al. 1997). The $\Delta a$ photometry is
from Vogt et al. (1998) with additional measurements as described in Section 
\ref{additional}. The $(g_1-y)$ data published by Vogt et al. (1998; see Table 4
therein) are not available in electronic form but were kindly provided by the authors. 

The following stars have inconsistent photometric measurements and were therefore
excluded: HR 345, HR 1375, HR 1617, HR 2870, HR 3470, HR 8854 and HR 8887. From the
available data, we are not able to decide if measurement errors, a wrong identification
or a binary nature causes these discrepancies. 

The final list comprises 225 objects that satisfy our criteria. The complete list
of program stars is only available in electronic form at the CDS via anonymous ftp to 
cdsarc.u-strasbg.fr (130.79.125.5), http://cdsweb.u-strasbg.fr/Abstract.html
or upon request from the first author. This table includes the
identification of objects, the $(g_{1}-y)_0$,
$(B-V)_0$, $X$ and $(u-b)$ values, $V$ magnitudes,
effective temperature with the corresponding errors (Fig. \ref{teff_plot} and
Sect. \ref{tcotet}), $v\sin i$ values and spectral types, respectively.

The distribution of Johnson $V$ and $(B-V)_0$ for our sample is shown in 
Fig. \ref{hists} (see Sect. \ref{reddening} for the estimation of the
reddening).
The peak of the $(B-V)_0$ values is at about $-$0.10\,mag (B8, Table 
\ref{standard}) which reflects the coincidence that the chemically 
peculiar stars which were measured within the $\Delta a$ photometric
system also peak at this effective temperature (Schneider 1993) and the
normal-type objects, used in this investigation, served as standard stars.

\subsection{Additional observations} \label{additional}

We observed a sub-sample of the bright B-type objects listed
in Cowley (1972) in the $\Delta a$ photometric system. 
She determined MK spectral types for all stars classified
as B8 in the Bright Star Catalogue north of $-$20$\degr$ in
order to investigate the appearance of CP objects. The results of
the $\Delta a$ photometry are discussed in Sect. \ref{deltaa}.

These photoelectric measurements in the $\Delta a$ photometric
system were carried out by one of us (A.~Schnell) at the 60\,cm
telescope of the Leopold-Figl-Observatory on Mitter-Sch{\"o}pfl
(880\,m above sea level, 40\,km southwest of Vienna). The telescope
was equipped with a single channel photometer and a thermoelectrically
cooled EMI~9844A photomultiplier using a diaphragm of 35$\arcsec$ (only
exceptionally 47$\arcsec$). The integration times per filter ranged from
15 to 25 seconds. The $g_1$ and $g_2$ filters are identical to
those described in Maitzen \& Floquet (1981) while the third one is a
conventional Str{\"o}mgren $y$ filter (Crawford 1978). The observations
were obtained over 26 nights between October 1991 and August 1996. 

The photometric measurements were reduced in the usual way, correcting for
airmass and extinction. The program stars were observed 
continuously over the years (up to 35 times)
which allows us to correct for different zero points and instrumental 
effects. The errors of the means for $a$ and $(g_1-y)$ are in the range 
between 1 and 7\,mmag, with a ``mean error'' of 3\,mmag only. This proves
the high quality of the individual measurements and supports
the reduction process.

Three stars (HR 1363, HR 1617 and HR 8821) are in common with the
paper by Vogt et al. (1998). The measurements of $(g_1-y)$ and $a$ agree
within 2\,mmag for these objects.

\begin{figure}
\begin{center}
\includegraphics[width=75mm]{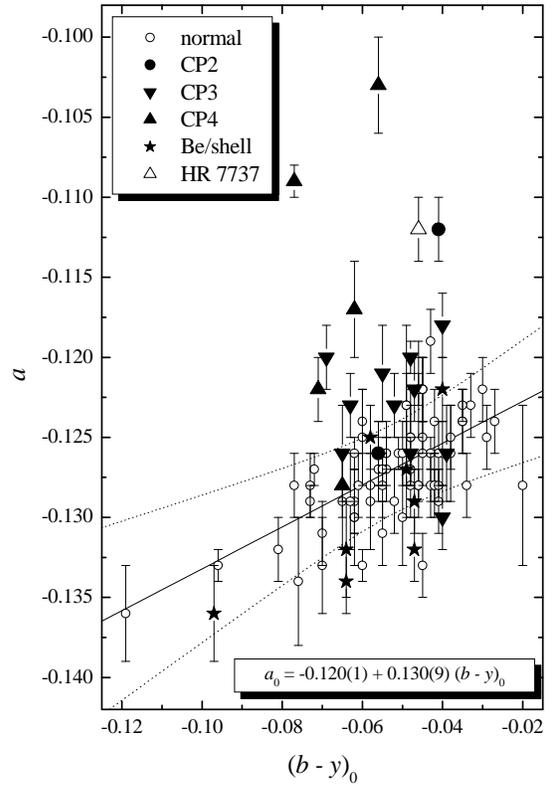}
\caption{Observed $a$ versus $(b-y)_0$
diagram for our sample of B-type objects. The solid line is the
normality line whereas the dotted lines are the confidence intervals
corresponding to 99.9\,\%.}
\label{da_plot}
\end{center}
\end{figure}

\subsection{$\Delta a$ photometry} \label{deltaa}

The $\Delta a$ intermediate band photometric system 
samples the depth of the 5200\AA\, flux depression 
by comparing the flux at the center ($\lambda_C$\,=\,5210\AA, 
FWHM\,=\,120\,\AA, $g_{\rm 2}$),
with the adjacent regions (5020\AA, 120\AA, $g_{\rm 1}$ and 5500\AA,
230\AA, $y$). The respective index was introduced as:
\begin{equation}
a = g_2 - (g_1 + y)/2
\end{equation}
It is optimized to detect magnetic chemically peculiar (CP2 and CP4) stars, 
but is also capable of detecting a certain percentage of 
non-magnetic CP1 and CP3 objects (Paunzen et al. 2005a). The nomenclature of
the different CP groups is according to Preston (1974). 
The group of classical Be/shell and metal-weak stars (Paunzen et
al. 2002) can be investigated. 
Since $a$ is dependant on the temperature, the intrinsic peculiarity index
had to be defined as
\begin{equation}
\Delta a = a - a_{\rm 0}[(b - y); (B - V); (g_{\rm 1} - y)]
\end{equation}
i.e. the difference between the individual $a$-values and those
of non-peculiar stars for the same color. The locus of the $a_{\rm 0}$-values
has been called the normality line.

Figure \ref{da_plot} shows the results of our observations
We indicate the different CP groups and known
Be/shell stars with individual symbols. The normality line was calculated 
using $(b-y)_0$ of all ``normal'' type objects, i.e. those not
included in the catalogue of peculiar objects by Renson et al. (1991):
\begin{equation}
a_0 = -0.120(1) + 0.130(9)\cdot(b-y)_0
\end{equation}
Known binary system were a-priori not excluded.
The 3\,$\sigma$ limit around the normality line is between 3 and 6\,mmag. 

As expected (Paunzen et al. 2005a),
the magnetic CP stars show the most significant positive $\Delta a$ values.
In general, CP3 stars tend to lie above the normality line whereas
Be/shell stars are below it. There are two exceptions, HR 6664 and HR 7721
which might be in their shell phases for which positive $\Delta a$ values
are expected (Maitzen \& Pavlovski 1987). We note that the two apparently
magnetic CP stars HR 481 (B8 IIIp Si) and HR 7401 (B8 IV He-weak)
show no significant positive $\Delta a$ values and should therefore be considered as
misclassified. 
Only one apparently normal type object, HR 7737 (B9 IV-V, $\Delta a$\,=\,+14\,mmmag),
deviates significantly from the normality line. This star is a known spectroscopic
binary system with unresolved components of 7.3 and 7.5\,mag (combined magnitude:
6.79\,mag; Germain et al. 1999). This seems to influence the photometric values
in the sense that it mimics a peculiar object.

\section{The estimation of the reddening} \label{reddening}

The reddening for B-type stars within the solar neighborhood is, in general, estimated
using photometric calibrations in the Str{\"o}mgren $uvby\beta$ 
(Crawford 1978) and the $Q$ parameter within the Johnson $UBV$ system (Johnson
1958). These methods are only based on photometric indices and do not take
into account any distance estimates via parallax measurements. 

\begin{figure}
\begin{center}
\includegraphics[width=65mm]{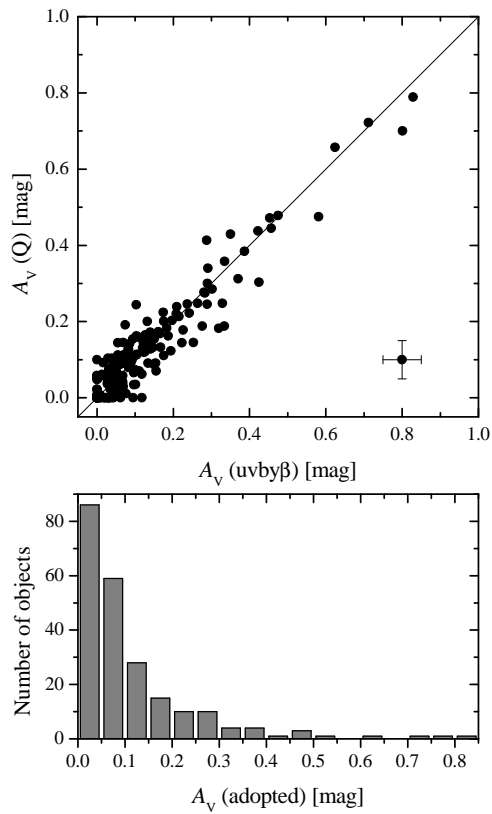}
\caption{The correlation of the absorption $A_V$ estimated from the
Str{\"o}mgren $uvby\beta$ and the $Q$ method (upper panel). The statistical
error for both methods is indicated with the symbol in the lower left
corner. The distribution of the adopted $A_V$ values is shown in the lower panel.}
\label{av_plot}
\end{center}
\end{figure}

The reddening in the Str{\"o}mgren $uvby\beta$ photometric system is based
on the comparison of the reddened $(b-y)$ and $c_1$ with the unreddened $(u-b)$
and $\beta$ indices (Crawford 1978). 
The procedure of the $Q$ method is straightforward and has been described
in detail by Johnson (1958). Here are the basic correlations from this
reference:
\begin{eqnarray}
Q &=& (U-B) - 0.72\cdot(B-V) -  \nonumber \\
&& 0.05\cdot(B-V)^2 \\
E(B-V) &=& (B-V) - 0.332\cdot Q
\end{eqnarray}
Throughout this paper, we use the following relation: 
$A_V$\,=\,3.1$E(B-V)$\,=\,4.3$E(b-y)$. Figure \ref{av_plot} shows the
comparison of the derived $A_V$ values for both methods. The agreement
is very good indicating that photometric measurements within the Johnson
and Str{\"o}mgren systems are consistent. Otherwise, a severe
deviation from the linear correlation would occur. The distribution of
the adopted values (mean of both methods) shows that most of the program
stars have an absorption which is below 0.1\,mag.

However, we also derived the interstellar reddening using the
model proposed by Chen et al. (1998) who combined galactic
reddening maps, which are derived from open clusters as well as from galactic
field stars with published empirical reddening laws from the literature. As 
input parameters, the galactic coordinates and the distance from the
Sun (i.e. derived from Hipparcos parallax measurements) are needed. The error
of the latter severely influences the error of the derived reddening. For
program stars with distance errors smaller than 15\%, a very good
agreement with the results of the photometric calibrations has been found. 

\section{The calibration of the effective temperature} \label{tcotet}

To reach our final goal, an empirical effective temperature calibration
of $(g1-y)_0$ for B-type stars, we calibrated this astrophysical
parameter for our sample using the published calibrations in the
Geneva, Str{\"o}mgren and Johnson photometric systems. Those calibrations
are derived independently of each other which allows us to
detect possible inconsistencies due to, for example, spectroscopic binary 
nature. We will now discuss the applied calibrations in more detail.
\\
{\it Geneva system:} detailed calibrations were published by
Cramer (1984, 1999) and K{\"u}nzli et al (1997). They are all based on the
reddening-free parameters $X$ and $Y$ which are valid for spectral types
hotter than approximately A0. The results are therefore independent of the
estimation of $A_V$ for our program stars. \\
{\it Str{\"o}mgren system:} the most recent and widely
used reference is Napiwotzki et al. (1993). For stars hotter than 11000\,K, the
unreddened $[u-b]$ and for cooler B-type objects, the $a_0$ index are used to
calibrate the effective temperature. The latter is not reddening free. \\
{\it Johnson system:} we calculated the $Q$ values for
luminosity class III and V objects according to the Tables listed by 
Schmidt-Kaler (1982). The $(B-V)_0$ values for those luminosity classes 
were transformed into effective temperatures using the results by Code et 
al. (1976, Table 7). As final correlation we derived a polynomial fit for
dependence of the effective temperature on $Q$ as
\begin{equation}
\log T_{eff} = 3.983(5) - 0.31(3)\cdot Q + 0.33(3)\cdot Q^2
\end{equation}
which is valid for all B-type, luminosity class III to V objects. 
This relation has to be treated as an averaged statistical result.
If one uses the individual measurements from Code et al. (1976) for the
ten stars that are in the relevant effective temperature and luminosity regime
(Sect. \ref{sops}), the following relation is derived:
\begin{equation}
\log T_{eff} = 3.994(29) - 0.25(15)\cdot Q + 0.38(15)\cdot Q^2
\end{equation}
A comparison of these two relations give $\Delta T_{eff}$\,=\,+141(147)\,K.

The individual effective temperature values within each
photometric system were first checked
for their intrinsic consistency and then averaged. These final values
together with the standard deviations of the means are listed in the
electronically available table. No outliers in any photometric system were detected.

\begin{figure}
\begin{center}
\includegraphics[width=80mm]{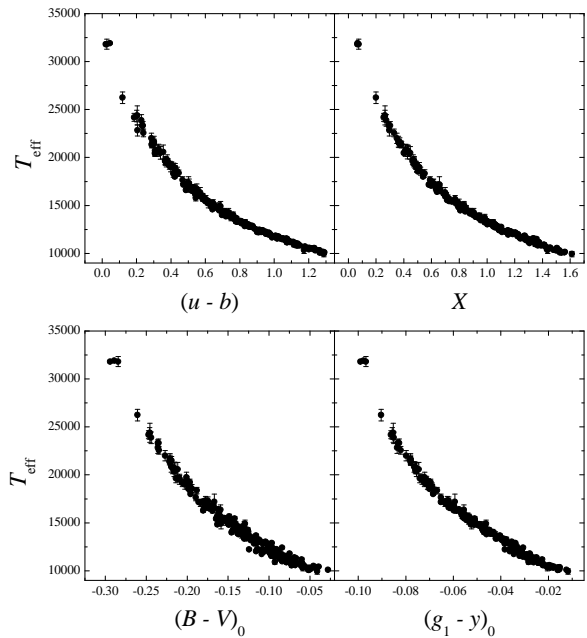}
\caption{Mean relation between the effective temperature and 
$(u-b)$, $X$, $(B-V)_0$ as well as $(g_1-y)_0$ for B-type, 
luminosity class V to III objects.}
\label{teff_plot}
\end{center}
\end{figure}

\section{The calibration procedure for the $\Delta a$ photometric system}

The starting point is one set of $(g_1-y)$ measurements, for example,
of an open cluster. The zero points
of data sets from different instruments and filter systems may vary. 
An overview of all currently used $\Delta a$ filters is listed in
Paunzen et al. (2005a, Table 2). They have also investigated the effects
on the absolute photometric values, which are in the range of the
observational errors for the photoelectric as well as CCD
technique (a few mmags only). 

The reddening coefficient $k$ of the relation
$E(g_1-y)$\,=\,$k \cdot E(B-V)$ was estimated which can then be
transformed to other photometric systems. We derived the
correlation coefficients for $(B-V)$ versus $(g_1-y)$ and then
applied these coefficients to $(B-V)_0$ using the estimated 
reddening for each individual star as well as $(g_1-y)$ to derive 
$(g_1-y)_0$ and thus $E(g_1-y)$. Our final value, $k$\,=\,0.39(2)
is in excellent agreement with the values derived from 
open cluster data (see Maitzen 1993 and references therein).

We then dereddened the individual indices and
established the linear correlations between those parameters.
We define a ``standard $(g_1-y)$ system'' which is set to
$(B-V)_0$\,=\,$(b-y)_0$\,=\,$(g_1-y)_0$\,=\,0 as follows:
\begin{eqnarray}
(B-V)_0 &=& -1.413(7) + 2.82(1)\cdot(g_1-y)_{0} \\
(b-y)_0 &=& -0.664(5) + 1.333(9)\cdot(g_1-y)_{0} 
\end{eqnarray}
This system will also be kept for the second paper.

Figure \ref{teff_plot} shows the relation between the effective
temperature and the different temperature sensitive indices 
for the four investigated photometric systems. The temperatures range
from approximately 33000 to 10000\,K which covers the 
main sequence B-type stars. However, there are only four stars 
with temperatures hotter than 25000\,K (B1.5\,V). We checked the
temperature relations with and without those four data points and found
no significant differences. Our final calibrations are:
\begin{eqnarray}
\log T_{eff} &=& + 4.520(3) - 0.754(15)\cdot(u-b) + \nonumber \\
&& + 0.413(2)\cdot(u-b)^2 - 0.108(11)\cdot(u-b)^3 \\
&=& + 4.546(3) - 0.698(13)\cdot X + \nonumber \\
&& + 0.367(17)\cdot X^2 - 0.092(6)\cdot X^3 \\
&=& + 3.956(5) - 1.04(7)\cdot(B-V)_0 + \nonumber \\
&& + 2.89(24)\cdot(B-V)_0^2 \\
&=& + 3.909(7) - 6.47(48)\cdot(g_1-y)_0 + \nonumber \\
&& - 47(9)\cdot(g_1-y)_0^2 - 425(60)\cdot(g_1-y)_0^3
\end{eqnarray}
with the mean of the errors for the whole sample of 
$\Delta T_{eff}[(u-b),X,(B-V)_0,(g_1-y)_0]$=[157,146,333,238\,K]. 
These are statistical errors and should be treated as such.
The mean relations between the effective temperature and $(B-V)_0$,
$(u-b)$, $X$ as well as $(g_1-y)_0$ depending on spectral types are
listed in Table \ref{standard}. The agreement with standard values 
for those indices from the literature (Code et al. 1976, Crawford 1978, 
Cramer 1999) is excellent.

\begin{figure}
\begin{center}
\includegraphics[width=80mm]{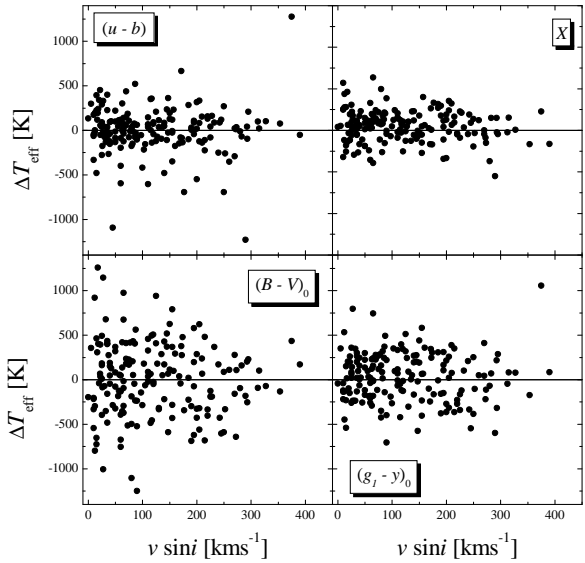}
\caption{No systematic correlation of $\Delta T_{eff}$ and the $v\sin i$
values were found for $(u-b)$, $X$, $(B-V)_0$ as well as $(g_1-y)_0$.}
\label{vsini}
\end{center}
\end{figure}

It is known that high rotational velocities can alter the photometric
indices significantly (Collins et al. 1991). The break-up velocity 
ranges from about 540 to 400\,kms$^{-1}$ for B0 to B9 stars (Townsend
et al. 2004). However, the inclination $i$ is the crucial point for 
the comparison of these models with observations.
The prototype A0 star Vega has a very low $v\sin i$ of
22\,kms$^{-1}$ but an equatorial velocity of 160\,kms$^{-1}$ (Hill
et al. 2004). We searched in the papers by Glebocki \& Stawikowski (2000)
and Abt et al. (2002) for available $v\sin i$ values for our program stars,
and averaged individual measurements from the different sources. In total,
there are measurements for 175 stars of our sample available with the highest
value of 390\,kms$^{-1}$ for HR~3502. Figure \ref{vsini} shows the diagrams
of $\Delta T_{eff}$\,=\,$T_{eff}(orig) - T_{eff}(calib)$ versus $v\sin i$ 
for the different indices. There is no obvious correlation between these two
parameters in any of the diagrams. The effect of the rotational
velocity is therefore less significant than the precision of the
method itself. Thus, the well-known alteration of 
stellar colors caused by high rotational velocities (Collins et al. 1991)
cannot be distinguished from the overall statistical errors resulting
from the calibration process.

The overall procedure of deriving the effective temperatures
for main sequence B-type objects in
the $\Delta a$ photometric system should be:
\begin{itemize}
\item Estimate the reddening, for example via isochrones for
open cluster members
\item $E(g_1-y)$\,=\,$0.39\cdot E(B-V)$
\item Apply the reddening correction for all individual indices
\item Transform the $(g_1-y)_0$ via the standard relations
\item Check the intrinsic consistency of all available measurements
according to the spectral type - effective temperature - photometric
indices correlation
\item Apply the effective temperature calibration
\end{itemize}
The estimated effective temperature from the $\Delta a$ system
can in addition be compared with calibrated values from other photometric indices.

\begin{table}[t]
\begin{center}
\caption{Mean relation between the effective temperature and $(B-V)_0$,
$(u-b)$, $X$ as well as $(g_1-y)_0$ for B-type, luminosity class V to III
objects.}
\label{standard}
\begin{tabular}{crcccc}
\hline
\hline
Spec. & $T_{eff}$ & $(B-V)_0$ & $(u-b)$ & $X$ & $(g_1-y)_0$ \\
\hline
B0 & 32000 & $-$0.292 & 0.020 & 0.060 & $-$0.102 \\
B1 & 26000 & $-$0.257 & 0.151 & 0.210 & $-$0.090 \\
B2 & 23000 & $-$0.236 & 0.239 & 0.310 & $-$0.082 \\
B3 & 18000 & $-$0.189 & 0.448 & 0.557 & $-$0.065 \\
B5 & 16000 & $-$0.164 & 0.570 & 0.705 & $-$0.056 \\
B6 & 14500 & $-$0.142 & 0.688 & 0.850 & $-$0.049 \\
B7 & 13500 & $-$0.125 & 0.785 & 0.970 & $-$0.043 \\
B8 & 12500 & $-$0.105 & 0.901 & 1.113 & $-$0.036 \\
B9 & 10800 & $-$0.064 & 1.155 & 1.411 & $-$0.021 \\
(A0) & 9800 & $-$0.031 & 1.338 & 1.603 & $-$0.011 \\
\hline
\end{tabular}
\end{center}
\end{table}

\section{Conclusion and outlook}

An empirical effective temperature calibration for the $\Delta a$
photometric system for main sequence (luminosity class V to III)
B-type stars was established. This system originally was developed
to search for magnetic chemically peculiar objects of the upper
main sequence. However, it also provides the temperature sensitive 
$(g_1-y)$ index as well as $y$ which can be used to determine the temperature
and absolute magnitude (or luminosity) of a star. For this purpose, data 
from the literature
as well as new observations were taken resulting in 225 suitably bright normal
type objects. Known binary stars and high $v\sin i$ stars were 
a-priori not excluded.
All program stars were first dereddened and then calibrated 
using already established methods in the Str{\"o}mgren $uvby\beta$, 
Geneva 7-color and Johnson $UBV$ photometric systems. These calibrations
are based on $(u-b)$, $X$ and $(B-V)_0$, respectively.

The intrinsic consistency for the individual stars was checked and
as final step, the averaged effective temperatures were calibrated in terms
of $(g_1-y)_0$ yielding a polynomial fit of third degree. The statistical
mean of the error for the sample of 225 stars is 238\,K which is very 
satisfactory and sufficient to investigate the HRD of distant galactic 
open clusters as well as extragalactic aggregates.

The second part of this series will investigate an empirical effective
temperature calibration of A-type to early F-type objects for which the calibration 
has to be more
sophisticated due to the increase of line blanketing and luminosity effects
without the availability of an independent reddening free parameter for the
Johnson $UBV$ and Geneva 7-color system.

\begin{acknowledgements}
This research was performed within the projects  
{\sl P17580} and {\sl P17920} of the Austrian Fonds zur F{\"o}rderung der 
wissen\-schaft\-lichen Forschung (FwF). 
It benefited from the financial contributions of the City of
Vienna (Hochschuljubil\"aumsstiftung project: Wiener Zweikanalphotometer).
Use was made of the SIMBAD database, operated at the CDS, Strasbourg, France and
of the NASA's Astrophysics Data System. We would like to thank Dr. Napiwotzki
for very useful comments and Dr. Kerschbaum
for providing photometric measurements electronically.
\end{acknowledgements}

\end{document}